\documentclass[onecolumn,showpacs,amsmath,amssymb,aps,showkeys,floatfix,prd,a4paper]{revtex4}

\usepackage[dvips]{graphicx}
\usepackage{dcolumn}
\usepackage{bm}
\usepackage{epsfig}
\usepackage{amsfonts}
\usepackage{amssymb,amscd}

\def\lsim{\raise0.3ex\hbox{$<$\kern-0.75em\raise-1.1ex\hbox{$\sim$}}}
\def\gsim{\raise0.3ex\hbox{$>$\kern-0.75em\raise-1.1ex\hbox{$\sim$}}}

\def\beq{\begin{equation}}
\def\eeq{\end{equation}}
\def\bea{\begin{eqnarray}}
\def\eea{\end{eqnarray}}
\def\bq{\begin{quote}}
\def\eq{\end{quote}}

\newcommand{\rqq}{\mbox{\boldmath $q$}}

\def\gappeq{\mathrel{\rlap {\raise.5ex\hbox{$>$}}
{\lower.5ex\hbox{$\sim$}}}}

\def\lappeq{\mathrel{\rlap{\raise.5ex\hbox{$<$}}
{\lower.5ex\hbox{$\sim$}}}}

\def\Toprel#1\over#2{\mathrel{\mathop{#2}\limits^{#1}}}

\newcommand{\rk}{\mbox{\boldmath $k$}}

\def\imag{{\mathcal{I}\mathrm{m}}\,}

\begin{document}


\title{Diffractive $J/\Psi$ photoproduction at large momentum transfer in  coherent hadron - hadron interactions at CERN LHC}

\author{V.~P. Gon\c{c}alves}
\email{barros@ufpel.edu.br}
\author{W.~K. Sauter}
\email{werner.sauter@ufpel.edu.br}
\affiliation{High and Medium Energy Group (GAME) \\
Instituto de F\'{\i}sica e Matem\'atica, Universidade Federal de Pelotas\\
Caixa Postal 354, CEP 96010-900, Pelotas, RS, Brazil}
\date{\today}

\begin{abstract}
The vector  meson production in coherent hadron-hadron interactions at LHC energies is studied  assuming  that the color singlet $t$-channel exchange carries large momentum transfer.  We consider the non-forward solution of the BFKL equation  at high energy and large momentum transfer and  estimate the rapidity distribution and total cross section for the process $h_1 h_2  \rightarrow h_1 J/\Psi X$, where $h_i$ can be a proton or a nucleus. We predict large rates,  which implies that the experimental identification can be feasible at the LHC.

\end{abstract}


\maketitle

\section{Introduction}
\label{intro}

Understanding the behavior of high energy hadron reactions from a fundamental perspective within of Quantum Chromodynamics (QCD) is an important goal of particle physics (For recent reviews see e.g. Ref. \cite{hdqcd,vera,iancu}). Attempts to test experimentally this sector of QCD started some years ago  with the first experimental results from $ep$ collisions at HERA and $pp (\bar{p})$ collisions at TEVATRON. Currently, there is a great expectation with respect to the first experimental results from hadron - hadron collisions at CERN LHC \cite{armesto}. 
{ The central papers concerning the knowledge of the Regge limit (high energy
limit) of QCD were presented} in the late 1970s by Lipatov and collaborators \cite{BFKL}. The physical effect that they describe is often referred to as the QCD Pomeron, with the evolution described by the BFKL equation.  One of main features of the BFKL evolution is the strong growth predicted for the cross sections, which implies that at very high energies the BFKL equation should be modified in order  to include unitarization corrections \cite{hdqcd,iancu}. However, it is { expected the existence of a} kinematical window in the current and future accelerators  in which the BFKL prediction should provide a good description of the experimental data (See e.g. \cite{bfkl_signatures}).

In recent years our group { has proposed} the analysis of coherent
interactions in hadronic collisions as an alternative way to study the
QCD dynamics at high energies
\cite{vicmag_upcs,vicmag_upcs2,vicmag_upcs3,vicmag_hq,vicmag_mesons_per,vicmag_prd,vicmag_pA,vicmag_difper,vicmag_mesons,vicmag_ane,vicmag_rho} (For related studies see \cite{klein_prc,Kleinpp,strikman_vec,motyka_watt,schafer1,cox}). The basic idea in coherent  hadronic collisions is that
the total cross section for a given process can be factorized in
terms of the equivalent flux of photons into the hadron projectile and
the photon-photon or photon-target production cross section.
The main advantage of using colliding hadrons and nuclear beams for
studying photon induced interactions is the high equivalent photon
energies and luminosities that can be { obtained} at existing and
future accelerators (For a review see Ref. \cite{upcs}).
Consequently, studies of $\gamma p (A)$ interactions
at the LHC could provide valuable information on the QCD dynamics.

The photon-hadron interactions in hadron-hadron collisions can be divided into exclusive and inclusive reactions. In the { former}, one given particle
is produced while the target remains in the ground state ({ or has only
internal excitations}), and, in the latter, the particle is produced
together with one or more particles { resulting} from the dissociation
of the target. The typical examples of these processes are the exclusive vector meson production and the inclusive heavy quark production,
described by the {reactions} $\gamma h \rightarrow V h$ ($V = \rho, J/\Psi, \Upsilon$) and $\gamma h \rightarrow X Y$ ($X = c\overline{c}, b\overline{b}$),
respectively. Recently, {both processes have been discussed considering} 
$pp$ \cite{vicmag_mesons_per,vicmag_prd,vicmag_mesons,vicmag_ane}, $pA$ \cite{vicmag_pA} and $AA$ \cite{vicmag_hq,vicmag_mesons_per,vicmag_mesons,vicmag_rho} collisions as an alternative to constrain the QCD dynamics at high energies (For reviews see Refs. \cite{vicmag_mpla,vicmag_jpg}), and the results demonstrate that their detection is feasible at the LHC.

\begin{figure}[t]
\centerline{\psfig{file=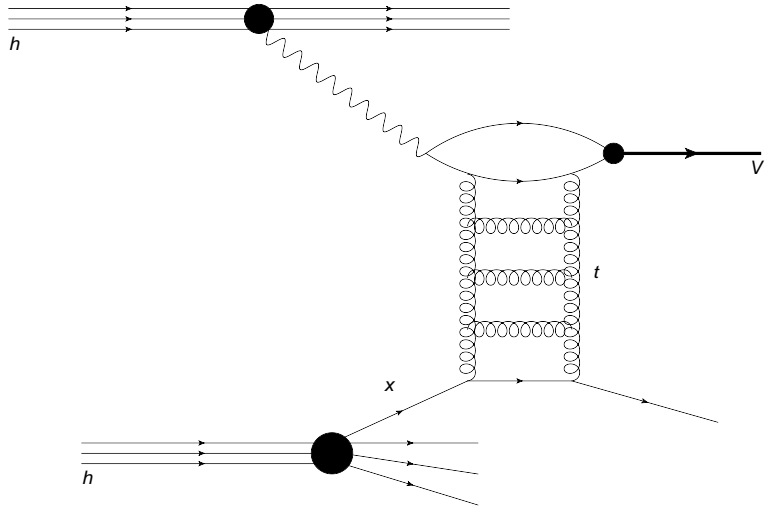,width=100mm}}
 \caption{High-$t$ vector meson photoproduction in coherent hadron - hadron collisions.}
\label{fig1}
\end{figure}

In this paper we extend the previous studies {to} the diffractive vector
meson photoproduction { with} hadron dissociation in the case of
large momentum transfer (See Fig. \ref{fig1}) and {estimate, for the first
time, the corresponding} cross section for $pp$ collisions at LHC (For related studies see Ref. \cite{fran}). In this process the { $t$-channel
color singlet} carries large momentum transfer, which  means that the square of the four momentum transferred {across} the associated rapidity gap, -$t$, is large. 
{Differently} from the diffractive processes studied in Refs. \cite{vicmag_mesons_per,vicmag_prd,vicmag_mesons,vicmag_rho,klein_prc,Kleinpp,strikman_vec,motyka_watt,schafer1,cox}, which are characterized by two rapidity gaps with the two hadrons remaining intact, now we still have  two  large rapidity gaps in the detector but  one of the hadrons dissociates. One expects a rapidity gap between the proton, which emits the photon and remains intact, and the vector meson.
{Concerning the other gap, we expect a vector meson on one side and a
jet on the other, which balances the transverse momentum}. {The present}
study is motivated by the fact that the experimental data for the $J/\Psi$ vector meson photoproduction at high $t$ in electron-proton collisions at HERA can be quite well described  using the impact factor representation and the  non-forward BFKL solution \cite{FP,jhep}.

This paper is organized as follows. In the next section we present a
brief review of the formalism necessary for calculate the vector meson production at high-$t$ in photon - hadron  and hadron - hadron collisions.  In Section \ref{results}  we present a comparison between  our predictions with the $ep$ HERA data for $J/\Psi$ production. Moreover, we present our predictions considering $pp$ and $AA$ collisions for LHC energies. Finally, in Section \ref{conc} we present a summary of our main conclusions.

\section{Formalism}
\label{form}

Let us consider the hadron-hadron interaction at large impact parameter ($b > R_{h_1}   R_{h_2}$) and ultra-relativistic energies. In this regime we expect  dominance of the  electromagnetic interaction.
In  heavy ion colliders, the heavy nuclei give rise to strong electromagnetic fields due to the coherent action of all protons in the nucleus, which can interact with each other. Similarly, this also occurs when considering ultra-relativistic  protons in $pp(\bar{p})$ colliders.
The photon emitted from the electromagnetic field
of one of the two colliding hadrons can interact with one photon of
the other hadron (two-photon process) or directly with the other hadron (photon-hadron
process). The total
cross section for a given process can be factorized in terms of the equivalent flux of photons of the hadron projectile and  the photon-photon or photon-target production cross section \cite{upcs}. In general,  the cross sections for $\gamma h$ interactions are two (or three!) orders of magnitude larger than for $\gamma \gamma$ interactions (See e.g. Ref. \cite{vicmag_upcs,Klein_vogt}). In what follows {we will focus on} photon - hadron processes. For a discussion of the double vector meson production at high-$t$ in $\gamma \gamma$ and ultra-peripheral heavy ion collisions see Refs. \cite{vicmag_upcs3,vic_werner,vic_werner2}.
Considering the requirement that in photoproduction there is no 
hadronic interaction (ultra-peripheral collision) an  analytical
approximation for the equivalent photon flux of a nucleus can be
calculated, and is given by \cite{upcs}
\begin{eqnarray}
\frac{dN_{\gamma}\,(\omega)}{d\omega}= \frac{2\,Z^2\alpha_{em}}{\pi\,\omega}\, \left[\bar{\eta}\,K_0\,(\bar{\eta})\, K_1\,(\bar{\eta})- \frac{\bar{\eta}^2}{2}\,{\cal{U}}(\bar{\eta}) \right]\,
\label{fluxint}
\end{eqnarray}
where
 $\omega$ is the photon energy,  $\gamma_L$ is the Lorentz boost  of a single beam; $K_0(\bar{\eta})$ and  $K_1(\bar{\eta})$ are the
modified Bessel functions.
Moreover, $\bar{\eta}=\omega\,(R_{h_1} + R_{h_2})/\gamma_L$ and  ${\cal{U}}(\bar{\eta}) = K_1^2\,(\bar{\eta})-  K_0^2\,(\bar{\eta})$.
{Eq. (\ref{fluxint})} will be used in our calculations of $J/\Psi$ photoproduction in  $AA$ collisions. For   proton-proton interactions we assume that the  photon spectrum is given by  \cite{Dress},
\begin{eqnarray}
\frac{dN_{\gamma}(\omega)}{d\omega} =  \frac{\alpha_{\mathrm{em}}}{2 \pi\, \omega} \left[ 1 + \left(1 -
\frac{2\,\omega}{\sqrt{S_\mathrm{NN}}}\right)^2 \right] 
\left( \ln{\Omega} - \frac{11}{6} + \frac{3}{\Omega}  - \frac{3}{2 \,\Omega^2} + \frac{1}{3 \,\Omega^3} \right) \,,
\label{eq:photon_spectrum}
\end{eqnarray}
with the notation $\Omega = 1 + [\,(0.71 \,\mathrm{GeV}^2)/Q_{\mathrm{min}}^2\,]$ and $Q_{\mathrm{min}}^2= \omega^2/[\,\gamma_L^2 \,(1-2\,\omega /\sqrt{S_\mathrm{NN}})\,] \approx (\omega/
\gamma_L)^2$.

The cross section for the diffractive $J/\Psi$ photoproduction at large momentum transfer  in a coherent  hadron-hadron collision is  given by
\begin{eqnarray}
\frac{d\sigma \,\left[h_1 + h_2 \rightarrow   h_1 \otimes J/\Psi \otimes X \right]}{dy dt} = \omega \frac{dN_{\gamma} (\omega )}{d\omega }\,\frac{d \sigma_{\gamma h \rightarrow J/\Psi  X}}{dt}\left(\omega \right)\,
\label{dsigdydt}
\end{eqnarray}
where $\otimes$ {means} the presence of a rapidity gap and the rapidity $y$ of the vector meson produced is directly related to the photon energy $\omega$, i.e. $y\propto \ln \, (2 \omega/m_{J/\Psi})$. Moreover, $\frac{d\sigma}{dt}$ is the differential cross section for the process $\gamma h \rightarrow J/\Psi X $. { Eq. (\ref{dsigdydt})} implies that, given the photon flux, the double differential cross {section} is a direct measure of the photoproduction
cross section for a given energy and squared momentum transfer. Some comments are in order here. {First}, the coherence condition {restricts} the photon virtuality to very low values, which implies that for most purposes, {the photons} can be considered as real. Moreover, if we consider
$pp/PbPb$ collisions at LHC, the Lorentz factor  is $\gamma_L = 7455/2930 $,
{which gives} the maximum {c.m.} $\gamma N$ energy $W_{\gamma p} \approx 8390/950$ GeV. Therefore, while studies of photoproduction at HERA are limited to photon-proton center of mass energies of about 200 GeV, photon-hadron interactions at  LHC can reach one order of magnitude higher on energy. Consequently, studies of coherent interactions at LHC could provide valuable information on the QCD dynamics at high energies.

The differential cross section $d \sigma /dt$  for the diffractive $J/\Psi$ photoproduction at large momentum transfer can be obtained using the impact  factor representation, proposed by Cheng and Wu \cite{ChengWu} many years ago (For a review see Ref. \cite{enberg}). In this representation, the amplitude for a large-$s$ hard collision process can be factorized in {three parts}: the impact factors of the colliding particles and the Green's function of two interacting reggeized gluons, which is determined by the BFKL equation and is  represented by ${\cal K}_\mathrm{BFKL}$ in what follows. The amplitude for the generic high energy process $A B\to C D$ can be expressed on the form 
\begin{equation}
{\cal A}^{A B\to C D} (\rqq) = 
\int {d^2 \rk \, d^2\rk^{\prime} }\,\,
 {\mathcal{I}}^{A\to C}(\rk, \rqq)\,\, \frac{{\cal{K}}_\mathrm{BFKL}(\rk,\rk^{\prime},\rqq)}{\rk ^2 (\rqq - \rk)^2} \,\,{\mathcal{I}}^{B\to D} (\rk^{\prime}, \rqq),
\label{mpr}
\end{equation}
where ${\mathcal{I}}^{A\to C}$ and ${\mathcal{I}}^{B\to D}$ are the impact factors for the upper and lower parts of the diagram, respectively.\ That is, they are the impact factors for the processes $A\to C$ and $B\to D$ with two gluons carrying transverse momenta $\rk$ and $\rqq-\rk$ attached. These gluons are in an overall color singlet state. At lowest order the process is described by two gluon exchange, which implies ${\cal K}_\mathrm{BFKL} \propto \delta^{(2)} (\rk - \rk^{\prime}) $ and an energy independent cross section. At higher order, the dominant contribution is given by the QCD pomeron singularity which is generated by the ladder diagrams with the (reggeized) gluon exchange along the ladder. The QCD pomeron is described by the BFKL equation \cite{BFKL}, with the exchange of a gluon ladder with interacting gluons generating a cross sections which increases with the energy.  As our goal  is the analysis of the vector meson production at large $-t$, we will use in our calculations the non-forward solution of the BFKL equation in the leading logarithmic approximation { (LLA)}, obtained by Lipatov in Ref.\cite{Lipatov}.

As discussed in detail in Refs. \cite{FR,BFLW}, at large momentum transfer the pomeron couples predominantly to individual partons in the hadron. This implies that the cross section for the photon - hadron interaction can be expressed by the product of the parton level cross section and the parton distribution of the hadron,
\begin{eqnarray}
\frac{d\sigma (\gamma h \rightarrow V X)}{dt dx_j} = \left[ \frac{81}{16} G(x_j,|t|) + \sum_j ( q_j(x_j,|t|) + \bar{q}_j(x_j,|t|))\right] \, \frac{d\sigma}{dt}(\gamma q \rightarrow V q)\,\,,
\label{dsigdtdx}
\end{eqnarray}
where $G(x_j,|t|)$ and $q_j(x_j,|t|)$ are the gluon and quark distribution functions, respectively.  
The struck parton initiates a jet and carries a fraction $x_j$ of the longitudinal momentum of the incoming hadron, which is given by $x_j = -t/(-t + M_X^2 - m^2)$, where $M_X$ is the mass of the {products of the target dissociation} and $m$ is the mass of the target. The minimum value of $x_j$ is calculated considering the experimental cuts on $M_X$.  
Following Refs. \cite{FP,jhep} we calculate $d \sigma /dt$ for the {process} $\gamma h \rightarrow V X$ {by} integrating Eq. (\ref{dsigdtdx}) over $x_j$ in the region $0.01 < x_j < 1$. 
The differential cross-section for the $\gamma q \rightarrow J/\Psi q$ process, characterized by the invariant collision energy squared $s$ of the photon - hadron system, is expressed in terms of the amplitude  ${\mathcal{A}}(s,t)$
as follows
\begin{equation}
\frac{d \sigma}{dt}(\gamma q \rightarrow V q) = \frac{1}{16 \pi} |{\mathcal{A}}(s,t)|^2.
\label{dsdt}
\end{equation}
The amplitude is dominated by its imaginary part, which we shall parametrize, as in \cite{FP,FR}, {through} a dimensionless quantity  $\mathcal{F}$
\begin{equation}
\imag {\mathcal{A}}(s,t) =
\frac{16 \pi}
{9 t^2} {\mathcal{F}}(z,\tau),
\label{FA}
\end{equation}
where $z$ and $\tau$ are defined by
\begin{equation}
z = \frac{3\alpha_{s}}{2\pi} \ln \biggl( \frac{ s}{\Lambda^{2}} \biggr),
\label{zdef}
\end{equation}
\begin{equation}
\tau = \frac{|t|}{M_{V}^{2}+ Q_{\gamma}^{2}},
\label{taudef}
\end{equation}
where $M_{V}$ is the mass of the vector meson, $Q_\gamma$ is the photon virtuality and $\Lambda^{2}$ is a characteristic  scale related to $M_V^2$  and $|t|$. In this paper we only consider $Q_\gamma=0$. In LLA, \( \Lambda  \) is arbitrary (but must depend on the scale in the problem, see discussion below) and \( \alpha _{s} \) is a constant. For completeness, we give the cross-section expressed in terms of ${\mathcal{F}}(z,\tau)$, where the real part of the amplitude is neglected,
\begin{equation}
\frac{d\sigma (\gamma q \rightarrow J/\Psi q)}{dt} \; = \;
\frac{16\pi}{81 t^4}
|{\mathcal{F}}(z,\tau)|^{2}.
\label{dsdtgq}
\end{equation}
This representation is rather convenient for the calculations performed below.

The BFKL amplitude, in the {LLA} and lowest conformal spin ($n=0$), is given by~\cite{Lipatov}
\begin{equation}
\label{BFKLa}
{\mathcal{F}}_{\mathrm{BFKL}}(z,\tau)=\frac{t^{2}}{(2 \pi)^{3}}\int d\nu \frac{\nu ^{2}}{(\nu ^{2}+1/4)^{2}}e^{\chi (\nu )z}I_{\nu }^{\gamma J/\Psi}(Q_{\perp })I^{q q}_{\nu }(Q_{\perp })^{\ast },
\end{equation}
 where $Q_{\perp}$ is the momentum transferred, $t=-Q_{\perp}^2$, (the subscript denotes {two-dimensional} transverse vectors) and 
\begin{equation}
\chi (\nu )=4{\mathcal{R}}\mathrm{e}\biggl (\psi (1)-\psi \bigg (\frac{1}{2}+i\nu \bigg )\biggr ) \label{eq:kernel}
\end{equation}
is proportional to the BFKL kernel eigenvalues~\cite{Jeff-book} with $\psi(x)$ being the digamma function.

\begin{figure}[t]
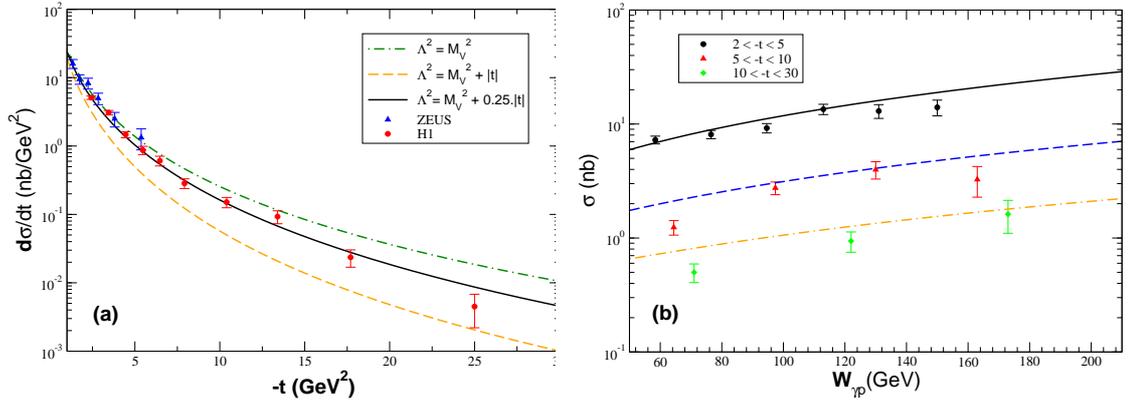

\begin{tabular}{cc}
\includegraphics[scale=0.3] {dsdtHERA.eps} & \includegraphics[scale=0.3]{stotHERA.eps}
\end{tabular}
\caption{(Color online) (a) Differential cross section for $J/\Psi$ production: theory compared to HERA data ($\langle W \rangle$ = 100 GeV).  (b) Energy dependence of the total cross section for distinct $t$-ranges. Data from H1 \cite{h1old} and ZEUS \cite{zeusold} Collaborations. }
\label{fig2}
\end{figure}

 The quantities $I_{\nu }^{\gamma J/\Psi}$ and $I_{\nu }^{qq}$ are given in terms of the impact factors ${\mathcal{I}}_{\gamma  J/\Psi}$ and ${\mathcal{I}}_{q q}$, respectively, and the BFKL eigenfunctions as follows \cite{FR}
\begin{eqnarray}
I_{\nu}^{ab}(Q_{\perp }) & = & \int \frac{d^{2}k_{\perp }}{(2\pi )^{2}}\, {\mathcal{I}}_{ab}(k_{\perp },Q_{\perp })\int d^{2}\rho _{1}d^{2}\rho _{2} \label{inugv}\\
 & \times  & \biggl [\left( \frac{(\rho _{1}-\rho _{2})^{2}}{\rho _{1}^{2}\rho _{2}^{2}}\right) ^{1/2+i\nu }-\left( \frac{1}{\rho _{1}^{2}}\right) ^{1/2+i\nu }-\left( \frac{1}{\rho _{2}^{2}}\right) ^{1/2+i\nu }\biggr ]e^{ik_{\perp }\cdot \rho _{1}+i(Q_{\perp }-k_{\perp })\cdot \rho _{2}}.\nonumber 
\end{eqnarray}
In the case of coupling to a colorless state only the first term in the square bracket { remains} since \( {\mathcal{I}}_{ab}(k_{\perp },Q_{\perp }=k_{\perp })={\mathcal{I}}_{ab}(k_{\perp }=0,Q_{\perp })=0 \). 
The impact factor ${\mathcal{I}}_{\gamma J/\Psi}$ describes, in the high energy limit, the couplings of the external particle pair to the color singlet gluonic ladder. It {is} obtained in the perturbative QCD framework and we approximate them by the leading terms in the perturbative expansion \cite{Ryskin}:
\begin{eqnarray}
{\mathcal{I}}_{\gamma J/\Psi}\;=\;
\frac{{\mathcal C} \alpha_s}{2}\, \biggl(\frac{1}{\bar{q}^{2}}-
\frac{1}{q_{\|}^{2}+k_{\perp}^{2}} \biggr).
\label{impfmom}
\end{eqnarray}
In this formula, {it is assumed the} factorization of the scattering process and the meson formation, and the non-relativistic approximation of the meson wave function is used. In this approximation the {quarks of the meson } have collinear four-momenta and $M_{J/\Psi}=2M_{c}$, where $M_{c}$ is the mass of the charm. To leading order accuracy, the constant ${\mathcal C}$ {can} be
related to the vector meson leptonic decay width
\begin{eqnarray}
\mathcal{C}^{2}\;=\;\frac{3\Gamma_{ee}^{J/\Psi}M_{J/\Psi}^{3}}{\alpha_{\mathrm{em}}}.
\end{eqnarray}
Moreover, we { have}
\begin{eqnarray}
\bar{q}^{2}\;=\;q_{\|}^{2}+Q_{\perp}^{2}/4, \\
q_{\|}^{2}\;=\;(Q_{\gamma}^{2}+M_{J/\Psi}^{2})/4.
\label{C}
\end{eqnarray}
Using Eq. (\ref{impfmom}) into (\ref{inugv}), one obtains \cite{FR,BFLW}
\begin{eqnarray}
I_{\nu}^{\gamma V_i}(Q_{\perp }) & = & -{\mathcal{C}_i}\, \alpha_s \frac{16\pi}{Q_{\perp }^{3}}\frac{\Gamma (1/2-i\nu )}{\Gamma (1/2+i\nu )}\biggl (\frac{Q_{\perp }^{2}}{4}\biggr )^{i\nu }\int _{1/2-i\infty }^{1/2+i\infty }\frac{du}{2\pi i}\biggl (\frac{Q_{\perp }^2}{4 M_{V_i}^2}\biggr )^{1/2+u}\\
 &  & \times\frac{\Gamma ^{2}(1/2+u)\Gamma (1/2-u/2-i\nu/2)\Gamma (1/2-u/2+i\nu/2)}{\Gamma (1/2+u/2-i\nu /2)\Gamma (1/2+u/2+i\nu /2)}.\nonumber \label{IV} 
\end{eqnarray}
The quark impact factor is given by ${\mathcal{I}}_{q q} = \alpha_s$, which implies \cite{BFLW}
 \begin{eqnarray}
I_{\nu}^{q q}(Q_{\perp }) & = & - \frac{4\pi \alpha_s}{Q_{\perp }} \left(\frac{Q_{\perp }^2}{4}\right)^{i\nu} \frac{\Gamma(\frac{1}{2}-i\nu)}{\Gamma(\frac{1}{2}+i\nu)}\,\,.
\label{iq}
\end{eqnarray}
The differential cross section can be directly calculated substituting the above expressions in Eq. (\ref{BFKLa}) and evaluating numerically the integrals. 

\begin{figure}[t]
\centerline{\psfig{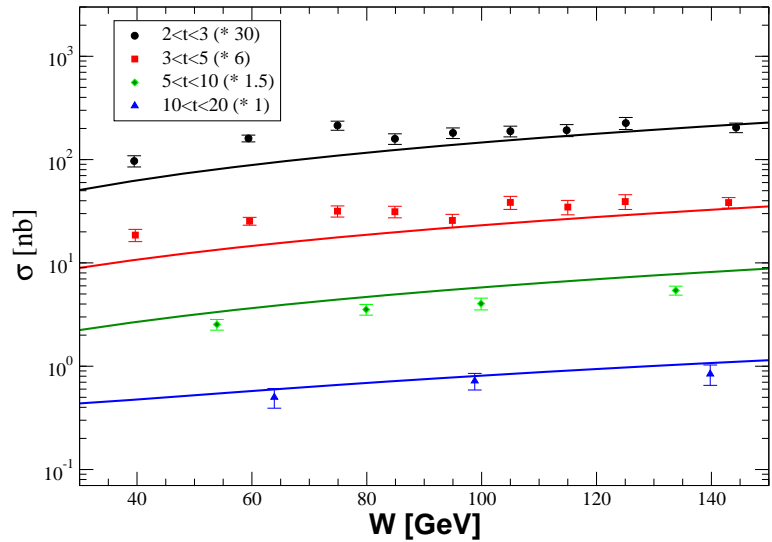}}
 \caption{(Color online) Energy dependence of the total cross section for distinct $t$-ranges. Data from ZEUS Collaboration \cite{zeusnew}. }
\label{fig3}
\end{figure}



\section{Results}
\label{results}

{Let us} start the analysis of our results discussing the diffractive $J/\Psi$ photoproduction in $ep$ collisions at HERA. In particular, it is important clarify our choice for the parameters $\alpha_s$ and $\Lambda$.
{The strong coupling appears in two pieces of calculations: in the
impact factors, coming from their couplings to the two gluons, and in the definition of the variable $z$ [Eq. (\ref{zdef})], being generated by the gluon coupling inside the gluon ladder} (For details see Ref.\cite{jhep}). In this paper we will treat these strong couplings as being identical and we will assume a fixed $\alpha_s$, which is appropriate to the leading logarithmic accuracy. Furthermore, as the cross section is proportional to $\alpha_s^4$, our results are strongly dependent on the choice {of} $\alpha_s$. { Following
Refs. \cite{FP,jhep}, we assume $\alpha_s = 0.21$, with this valued determined
from a fit to the HERA data} (For a detailed discussion see Ref. \cite{FP}).
Similarly, {in the LLA},  $\Lambda$  is arbitrary but must depend on the scale in the problem. In our case we have that in general $\Lambda$ will be a function of $M_V$ and/or $t$. Following Ref. \cite{FP} we {assume} that $\Lambda$ can be expressed by $\Lambda^2 = \beta M_{V}^2 + \gamma |t| $, with $\beta$ and $\gamma$ being free parameters to be fitted by data. 
In Fig. \ref{fig2} (a) we compare our predictions for the differential cross section for $J/\Psi$ production with the HERA data \cite{h1old,zeusold}. The curves were obtained using the CTEQ6L parton distributions \cite{cteq6} and considering distinct values for $\beta$ and $\gamma$. A very good agreement with the data is obtained for $\beta = 1.0$ and $\gamma = 0.25$, which are the values that we will use in what follows. In Fig. \ref{fig2} (b) we compare our predictions for the total cross section, obtained { by integrating} the differential cross sections over distinct $t$-ranges, with the HERA data. We can observe that the experimental data are quite well described, specially at larger values of the center-of-mass energy. Very recently, the ZEUS Collaboration has reported results \cite{zeusnew}  for the $J/\Psi$ photoproduction at large-$t$ in the kinematical range $30 < W < 160$ GeV and $2 < |t| < 20$ GeV$^2$, which is larger than {that in previous ZEUS measurements} \cite{zeusold}. In Fig. \ref{fig3} we compare our predictions with these new data. We have that our predictions describe the data at large-$t$ and/or large values of energy  but underestimate the data at small values of $t$ and $W$.

\begin{figure}[t]
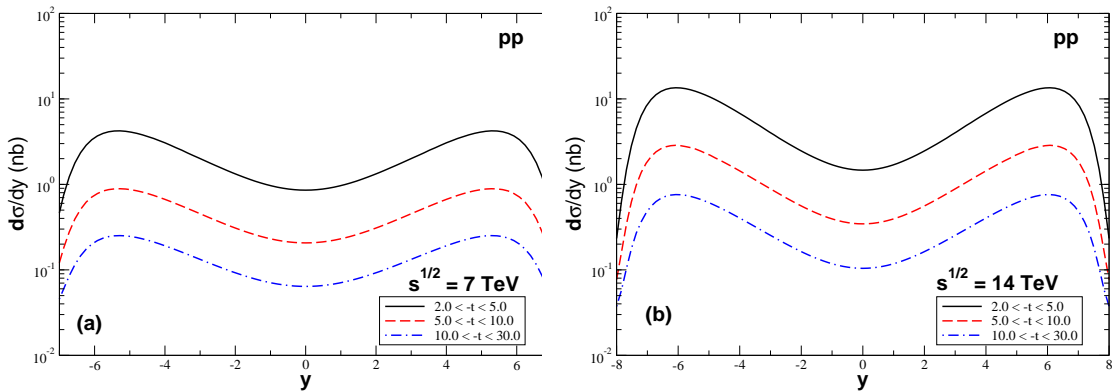

\begin{tabular}{cc}
\includegraphics[scale=0.3] {dsdypp.eps} & \includegraphics[scale=0.3]{dsdypp14.eps}
\end{tabular}
\caption{(Color online) Rapidity distribution for the diffractive $J/\Psi$ photoproduction in $pp$ collisions at LHC for distinct $t$-ranges and different values of the center-of-mass energy: (a) $\sqrt{s} = 7.0$ TeV and  (b) $\sqrt{s} = 14.0$ TeV. }
\label{fig4}
\end{figure}

\begin{figure}[t]
\centerline{\psfig{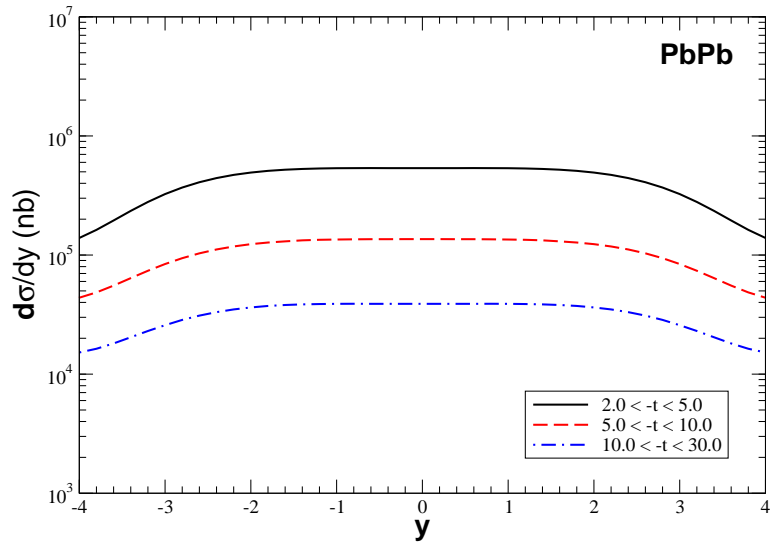}}
 \caption{(Color online) Rapidity distribution for the diffractive $J/\Psi$ photoproduction in $PbPb$ collisions at LHC for distinct $t$-ranges and  $\sqrt{s} = 5.5$ TeV. }
\label{fig5}
\end{figure}

{Let us} now calculate the rapidity distribution and total cross sections for diffractive $J/\Psi$ photoproduction in coherent hadron-hadron collisions. The distribution {in} rapidity $y$ of the produced final state can be directly computed from Eq. (\ref{dsigdydt}), by integrating over the squared transverse momentum and using the corresponding  photon spectrum.  We consider three different choices for the limits of integration. Basically, we assume the same values used in Ref. \cite{zeusold} by ZEUS Collaboration. This choice is directly associated to the fact that the associated $\gamma p$ data are quite {well described} by our  formalism (See Fig. \ref{fig2}). 
Moreover, we  integrate over $x_j$  in the range $ 10^{-2} < x_j < 1$, which {means} that we {assume} that the upper limit on the mass of the target dissociation products at LHC is similar to that considered at HERA. {Our
predictions} for the rapidity distribution increase by 20\% if  the minimum value of $x_j$ is assumed to be $10^{-3}$.
In Figs. \ref{fig4} (a) and (b) we present our results for the rapidity distribution considering $pp$ collisions at $\sqrt{s} = 7.0$ TeV and  $\sqrt{s} = 14.0$ TeV, respectively. In this case we consider that the photon spectrum is 
 described by Eq. (\ref{eq:photon_spectrum}).  As expected from Fig. \ref{fig1} (a), the rapidity distribution decreases when we select a range with larger values of $t$. Furthermore, it increases with the energy, which is directly associated to the LL BFKL dynamics which predicts a strong growth with the energy for the cross sections ($\sigma_{\gamma h} \propto W^{\lambda}$ with $\lambda \approx 1.4$). In Fig. \ref{fig5} we present our results for $PbPb$ collisions and $\sqrt{s} = 5.5$ TeV. In this case we assume that the photon spectrum is described by Eq. (\ref{fluxint}) and that the nuclear parton distributions are given by the EKS parametrization \cite{eks}.  In comparison to the $pp$ collisions, for heavy ion interactions we predict very {larger} values for  the rapidity distribution with a similar $t$-dependence. Moreover,  we predict a plateau in the range $|y| < 2 $ .



In Table \ref{tab1} we present our predictions for the  total cross section for some values of center of mass energy considering $pp$ and $PbPb$ collisions. 
The  BFKL dynamics implies that the cross sections strongly increases with the energy, resulting in an enhancement {by} a factor of about 3 when the energy is increased from 7 to 14 TeV. Moreover, {since} the photon flux is proportional to $Z^2$,
{because} the electromagnetic field surrounding the ion is very larger than the proton one due to the coherent action of all protons in the nucleus,  the nuclear cross sections are amplified by a factor $Z^4$, which implies very large cross sections for the diffractive $J/\Psi$ photoproduction at large-$t$ in $PbPb$ collisions at LHC.
Considering the {design} luminosities at LHC for $pp$ collisions (${\cal L}_{\mathrm{pp}} = 10^{34}$ cm$^{-2}$s$^{-1}$) and $PbPb$ collisions (${\cal L}_{\mathrm{PbPb}} = 4.2 \times 10^{26}$ cm$^{-2}$s$^{-1}$) we can calculate the production rates (See Table \ref{tab1}). Although the cross section for the diffractive $J/\Psi$ photoproduction at large-$t$ in $AA$ collisions is much larger than in $pp$ collisions, the event rates are higher in the $pp$ mode  due to its larger luminosity. In particular, we predict approximately 970 events per second in the range $2.0 < |t| < 5.0$ for $pp$ collisions at $\sqrt{s} = 14$ TeV. In contrast, 13 events per second are predicted for $PbPb$ collisions at $\sqrt{s} = 5.5$ TeV in the same $t$-range. However, 
for a luminosity above ${\cal L} \ge 10^{33}$ cm$^{-2}$s$^{-1}$, multiple hadron - hadron collisions per bunch crossing are { very likely}, which leads to a relatively large occupancy of the detector channels even at low luminosities.
{This} drastically reduces the possibility of {measurement} of coherent
processes at these luminosities. In contrast, at lower luminosities the event pile-up is negligible. Consequently, an estimative considering ${\cal L}_{\mathrm{pp}} = 10^{32}$ cm$^{-2}$s$^{-1}$ should be more realistic. It reduces our predictions for the event rates in $pp$ collisions by a factor $10^2$.

We predict  very large cross sections, which implies that
{it would be possible to collect an impressive statistics provided one can
devise} an effective trigger for the two low transverse momentum leptons not accompanied by {hadron production}.
This is a current  experimental challenge. Simulations indicate that these events can be identified with a good signal to background ratios when the entire event is reconstructed and a cut is applied on the summed transverse momentum of the event \cite{david}. 
Another important aspect that restricts the experimental separation of coherent processes are the basic features of the LHC detectors \cite{cms,f420,alice,atlas}. The coherent condition { leads to the result
that} the final state has a very low transverse momentum. Consequently, the decay 
{electrons or muons from $J/\Psi$} are produced basically at rest and have energies of {the order} of $m_{J/\Psi}/2$. Such energies are  in a range close to the lower limits of detector acceptance and trigger selection thresholds. In particular, the energy of the decay products of $J/\Psi$  is lower than those needed to reach the electromagnetic calorimeter or muon chambers at CMS and ATLAS without significant energy losses in the intermediate material. Probably,  the detection of the $J/\Psi$ produced in coherent processes will be not possible  in the CMS and ATLAS detectors, and only the $\Upsilon$ states could be observed \cite{cms}. 
In Ref. \cite{cox} the authors have estimated the effect of detector acceptances and the impact of measuring one of the protons in the diffractive $\Upsilon$  photoproduction in $pp$ collisions. In particular, they have estimated the effect of the inclusion of a cut in the angle of emission and in the transverse momentum of the decay dileptons. Moreover, the possibility to detect one of the protons in region 420 m from the interaction point was analysed. This study concludes that these cuts diminishe   the cross  section for $\Upsilon$ production. However,  it still is feasible to constrain the QCD dynamics using this process.
The implication of these cuts for $J/\Psi$ production, taking into account the characteristics  of the CMS and ATLAS detectors, is a subject that deserves a more detailed study. We postpone this analysis for a future publication. However, it is important to emphasize that in ALICE, which is designed to handle multiplicities of several thousand particles in a single event, the reconstruction of the low {multiplicity} events associated to coherent processes should not be a problem \cite{alice}.  In particular, the muon arm, which covers the pseudorapidity range $-2.5 > \eta > -4.0$,   should be  capable of reconstructing $J/\Psi$ and $\Upsilon$  vector mesons through their dilepton decay channel.



\begin{table}[t]
\begin{center}
\begin{tabular}{||c|r@{.}l|r@{.}l|r@{.}l||}
\hline
\hline
  & \multicolumn{2}{c|}{$pp$ ($\sqrt{s} = 7$ TeV)} & \multicolumn{2}{c|}{$pp$ ($\sqrt{s} = 14$ TeV)} & \multicolumn{2}{c||}{$PbPb$ ($\sqrt{s} = 5.5$ TeV)}  \\
\hline
$2.0 < |t| < 5.0$  & 32 &0 nb (320.0) & 97&0 nb (970.0) & 3&0 mb (13.0)  \\
\hline
$5.0 < |t| < 10.0$  & 7 &0 nb (70.0)& 21&0 nb (210.0) & 0&9 mb  (0.38) \\
\hline
$10.0 < |t| < 30.0$  & 2  &0 nb (20.0) & 6&0 nb (60.0)& 0&3 mb  (0.12)\\
\hline
\hline
\end{tabular}
\end{center}
\caption{The integrated cross section (event rates/second)  for the diffractive $J/\Psi$ photoproduction at large momentum transfer in $pp$ and $AA$ collisions at LHC.} 
\label{tab1}
\end{table}

{ Let us} compare our results with those presented in Ref. \cite{vicmag_mesons} where  the diffractive $J/\Psi$ photoproduction at $t = 0$ in coherent interactions was calculated considering the Color Glass Condensate formalism \cite{hdqcd}. Both processes predict the presence of two rapidity gaps. Our predictions are {smaller than those shown in \cite{vicmag_mesons}
at least by a factor 1.5}. Another background which is characterized by two rapidity gaps in the final state are diffractive hadron - hadron interactions: $h_1 + h_2 \rightarrow h_1 \otimes V \otimes h_2$. Recently, the exclusive $J/\Psi$ and $\Upsilon$ hadroproduction in $pp/p\bar{p}$ collisions were estimated in Ref. \cite{motyka} considering the pomeron-odderon fusion (See also \cite{schafer}). Although there is a large uncertainty in the predictions for pomeron-odderon fusion, it can be { of the order} of our predictions for the photoproduction of vector mesons. As pointed in Ref. \cite{motyka}, the separation of odderon and photon contributions should be feasible by the analysis of the outgoing momenta distribution. Furthermore, it is important to emphasize that the experimental separation from the process studied in this paper should be possible, since that differently from the processes discussed above, at large momentum transfer one of the projectiles  dissociates. Therefore, this process could be separated using  the forward detectors proposed to be installed  at LHC \cite{forward}.

Finally, we would like to emphasize that several points in the present calculation deserve more detailed studies: the contribution of all conformal spins \cite{BFLW,jhep,spin}, the helicity flip  of quarks \cite{jhep}, the next-to-leading order (NLO) corrections to the BFKL dynamics \cite{nlo_bfkl} and the corrections associated to the saturation effects \cite{hdqcd,iancu}. From previous studies ~\cite{jhep}, {one expects} the contribution of higher conformal spins to be small. However, the NLO corrections should modify the energy dependence of the $\gamma h$ cross section and consequently our predictions for the rapidity distributions and total cross sections. In order to estimate these modifications it is necessary to include the corrections {to} the BFKL Pomeron, as {it was done} e.g. in  \cite{Brodsky:1998kn}, and for the $\gamma \rightarrow V$ impact transition factor ~\cite{Ivanov:2004pp}.
{We intend to take into account these contributions in future publications}.


\section{Summary}
\label{conc}

The LHC offers a unique possibility to probe QCD in a new and hitherto unexplored regime. In particular, it will allow { the study of} coherent processes which are characterized by photon - hadron and photon - photon interactions. In this paper we { have restricted} our study {to} $\gamma h$ interactions and extended {previous analysis to} the diffractive $J/\Psi$ photoproduction { together with} hadron dissociation {with} large momentum transfer. The rapidity distribution and total  cross sections were estimated considering distinct center-of-mass energies and projectiles using  the non-forward  solution of the BFKL equation  at high energy and large momentum transfer.
Our main conclusion is that the LHC can experimentally {check our}
predictions. {Besides,} we believe that this process can be used to constrain the QCD dynamics at high energies. 

\section*{Acknowledgements}
 VPG would like to thanks J.T. de Santana Amaral by useful discussions. This work was partially financed by the Brazilian funding agencies CNPq and FAPERGS.



\end{document}